\documentclass[onecolumn,showpacs,aps,floatfix,pra,superscriptaddress,nofootinbib]{revtex4}
\usepackage{color}
\usepackage{graphicx}
\usepackage{bm}
\usepackage{amsmath}
\begin{document}

\title{Strong quantum violation of the gravitational weak equivalence principle
 by a non-Gaussian wave-packet}
\author{P. Chowdhury}
\email{priyanka@bose.res.in}
\affiliation{S. N. Bose National Center
for Basic Sciences, Block JD, Sector III, Salt Lake, Kolkata 700 098, India}
\author{D. Home}
\email{dhome@bosemain.boseinst.ac.in} 
\affiliation{CAPSS, Department
of Physics, Bose Institute, Sector-V, Salt Lake, Kolkata 700 091,
India}
\author{A. S. Majumdar}
\email{archan@bose.res.in} 
\affiliation{S. N. Bose National Center
for Basic Sciences, Block JD, Sector III, Salt Lake, Kolkata 700
098, India}
\author{S. V. Mousavi}
\email{vmousavi@qom.ac.ir} 
\affiliation{Department of Physics, 
The University of Qom, Qom 3716146611, Iran}

\affiliation{Institute for studies in Theoretical Physics and Mathematics (IPM), P.O. Box 19395-5531, Tehran, Iran}
\author{M. R. Mozaffari}
\email{m.mozaffari@qom.ac.ir} 
\affiliation{Department of Physics,
The University of Qom, Qom 3716146611, Iran}
\author{S. Sinha}
\email{siddhartha@bose.res.in}
\affiliation{S. N. Bose National Center
for Basic Sciences, Block JD, Sector III, Salt Lake, Kolkata 700
098, India}
\begin{abstract}
The weak equivalence principle of gravity is examined at the quantum level in 
two ways. First, the position detection probabilities of particles described 
by a non-Gaussian wave-packet projected upwards against gravity around the 
classical turning point and also around the point of initial
projection are calculated. These probabilities exhibit mass-dependence at both 
these points, thereby reflecting the quantum violation of the weak equivalence 
principle. Secondly, the mean arrival time of freely falling particles is 
calculated using the quantum probability current, which also turns out to be 
mass dependent. Such a mass-dependence is shown to be enhanced by increasing 
the non-Gaussianity parameter of the wave packet, thus signifying a stronger
violation of the weak equivalence principle through a greater departure from
Gaussianity of the initial wave packet. The mass-dependence of both the 
position detection probabilities and the mean arrival time vanish in the limit 
of large mass. Thus, compatibility between the weak equivalence principle and 
quantum mechanics is recovered in the macroscopic limit of the latter. A 
selection of Bohm trajectories is exhibited to illustrate these features in 
the free fall case. 
\end{abstract}

\maketitle

%NewSection-NewSection-NewSection-NewSection-NewSection-NewSection-NewSection-
\section{Introduction}

\vspace{1cm}
\hspace{0.5cm}
In the famous gedanken experiment conceived by Galileo, the universality of the ratio between the gravitational and inertial masses has been studied with test bodies in free fall from the leaning tower of Pisa \cite{Ga-book-1638}. Since then several tests of weak equivalence principle have been performed with classical test bodies such as very sensitive pendula or torsion balances establishing extremely accurate conformation of the equality of gravitational and inertial masses. The most familiar tests of the weak equivalence principle are experiments of the E\"{o}tv\"{o}s-type \cite{Eo-Math-1890}, which measure the gravitational acceleration of macroscopic objects.Thus Einstein's weak equivalence principle is established in the Newtonian picture of the gravitational force which is exactly proportional to the inertial mass. The validity of the principle even for quantum mechanical particles is also proved using gravity-induced interference experiments \cite{CoOvWe-PRL-1975, PeChCh-Nature-1999}.\\

The traditional equivalence principle is fundamentally both classical and local, and it is interesting to enquire how it is to be understood in quantum mechanics. Principle of equivalence states that inertial mass is equal to gravitational mass; $ m_i=m_g=m$. Another statement is that when all sufficiently small test bodies fall freely, they acquire an equal acceleration independent of their mass or constituent in a gravitational field. Quantum mechanically, it \cite{Ho-book-1993} can be said that " The results of experiments in an external potential comprising just a sufficiently weak, homogeneous gravitational field, as determined by the wave function, are independent of the mass of the system". The last statement is also known as weak equivalence principle of quantum mechanics (WEQ).\\

The possibility of violation of weak equivalence principle in quantum mechanics is discussed in a number of papers, for instance using neutrino mass oscillations in a gravitational potential \cite{Ga-PRD-1988}. The evidence of violation of WEQ can be shown both experimentally and theoretically. Experimental evidence exists in the interference phenomenon associated with the gravitational potential in neutron and atomic interferometry experiments \cite{CoOvWe-PRL-1975, PeChCh-Nature-1999}. Theoretically, for a particle bound in an external gravitational potential, it is seen that the radii, frequencies and binding energy depend on the mass of the bound particle \cite{Gr-RMP-1979, Gr-AP-1968, Gr-RMP-1983}.\\

The free fall of test particles in a uniform gravitational field in the case of quantum states with and without a classical analogue was re-examined by  Viola and Onofrio\cite{Vi-PRD-1997}. Another quantum mechanical approach of the violation of WEQ was given by Davies \cite{Da-CQG-2004} for a quantum particle in a uniform gravitational field using a model quantum clock \cite{Pe-AJP-1980}. A quantum particle moving in a gravitational field can penetrate the classically forbidden region beyond the classical turning point of the gravitational potential and the tunnelling depth depends on the mass of the particle. So, there is a small mass-dependent quantum delay in the return time representing a violation of weak equivalence principle. This violation is found within a distance of roughly one de Broglie wavelength from the turning point of the classical trajectory.\\

In the recent past, Ali et al. \cite{Al-CQG-2006} have shown an explicit mass dependence of the position probabilities for quantum particles projected upwards against gravity around both the classical turning point and the point of initial projection using Gaussian wave packet. They also have shown an explicit mass dependence of the mean arrival time at an arbitrary detector location for a Gaussian wave packet under free fall. Thus both the position probabilities and the mean arrival time show the violation of WEQ for quantum particles represented by Gaussian wave packets. In the present paper, using a class of non-Gaussian wave packets involving a tunable parameter, we shall compute position detection probabilities around the classical turning point and the point of initial projection and also compute the mean arrival time for atomic and molecular mass particles represented by non-Gaussian wave packets to observe stronger violation of WEQ for quantum particles. To compute the mean arrival time, we consider the case when the particles are dropped from a height with zero initial velocity.\\

Violation of WEQ arises as a consequence of the spread of wave packets the 
magnitude of which itself depends on the mass. In order to illustrate this
effect we present an analysis in terms of Bohmian trajectories. In the Bohmian model \cite{Bo-PR-1952}  each individual particle is assumed to have a definite position, irrespective of any measurement. The pre-measured value of position is revealed when an actual measurement is done. Over an ensemble of particles having the same wave function $\psi$, these ontological positions are distributed according to the probability density $\rho=|\psi|^2$ where the wave function $\psi$ evolves with time according to the Schr\"odinger equation and the equation of motion of any individual particle is determined by the guidance equation $v = J/\rho$, where $v$ is the Bohmian velocity of the particle and $J$ is the probability current density. Solving the guidance equation one gets the trajectory of the particle. In the present work we describe the evolution of Bohmian trajectories
in terms of the mass and the tunable parameter of the non-Gaussian wave packet.

%NewSection-NewSection-NewSection-NewSection-NewSection-NewSection-NewSection-

\section{Position Detection Probability}
\vspace{1cm}
\hspace{0.5cm}

Let us consider an ensemble of quantum particles projected upwards against 
gravity with a given initial mean position and mean velocity. Our motivation 
is to look for greater violation of the equivalence principle compared to that
achieved by a Gaussian wave packet. Hence, we formulate a one-dimensional 
non-Gaussian wave function which represents the initial state of each particle, 
given by
\begin{equation}
\psi(z,t=0)=N \left[ 1+\alpha \sin{ \left( \frac{\pi}{4\sigma_0} z \right) } \right] e^{-\frac{z^{2}}{4\sigma_0^2}+ik_{0}z}
\end{equation}
\label{iwf}
Here, $ \alpha $ is the tuneable parameter varying from $0$ to $1$. The normalisation factor $ N $ is given by
\begin{eqnarray}
N &=& \left( \sqrt{2\pi} \sigma_0 \left[ 1+\frac{\alpha^{2}}{2}\:( 1-e^{-\pi^{2}/8})\right] \right)^{-1/2}~.
\end{eqnarray}
The salient features corresponding to the above wave packet are its asymmetry,
its infinite tail, and its reduciblity to a Gaussian wave packet upon a 
continuous decrement of the parameter $\alpha$ to zero.  The property of
an infinite tail of the above wave function is not associated with
non-Gaussian forms that are generated by truncating the Gaussian distribution.
The asymmetry of the wave packet (due to the sine function) entails a 
difference between its mean and peak, as will be further evident in our
following analysis.

The initial group velocity is given by $ u=\frac{\hbar k_0}{m} $, where $ m $ is the mass of the particle and $ k_{0} $ is the wave number. For $ \alpha =0 $, we get the Gaussian wave function.
If $ \psi_{1} $ and $ \psi_{2} $ represent the initial wave functions for particles 1 and 2 in the Schr\"{o}dinger picture, then the mean initial conditions are
\begin{eqnarray}
\langle \hat{z}\rangle_{\psi_{1}}=\langle \hat{z}\rangle_{\psi_{2}} = \langle \hat{z}(0)\rangle \nonumber\;,\\
\frac{\langle\hat{p}_{z}\rangle_{\psi_{1}}}{m_{1}}=\frac{\langle\hat{p}_{z}\rangle_{\psi_{2}}}{m_{2}}=u
\end{eqnarray}
\label{initialcondt}
where $ \langle \hat{z}\rangle_{\psi} $ and $ \langle\hat{p}_{z}\rangle_{\psi} $ represent the expectation values for position and momentum operators respectively. Here, we have restricted ourselves to a one-dimensional representation along the vertical $ z $ direction for simplicity.
With the propagator
\begin{equation}
G(z,t|y,0)=\sqrt{\frac{m}{2\pi i\hbar t}} e^{\frac{im}{2\hbar t}(z-y)^{2}-\frac{imgt}{2\hbar}(z+y)-\frac{img^2t^{3}}{24\hbar}}
\end{equation}
of a particle in the linear gravitational potential $ V = mgz $ and with the relation
\begin{equation}
\psi(z,t)=\int dy G(z,t|y,0)\psi(y,0)
\end{equation}
one can get the Schr\"{o}dinger time evolved wave function $ \psi(z,t) $ at any subsequent time $t$ as

\begin{eqnarray}\label{eq: psit}
\psi(z,t)&=& N \sqrt{\frac{\sigma_0}{s_t}}~e^{\frac{im}{2\hbar t}
[z^2 - g t^2 z - \frac{g^2 t^4}{12}]} \nonumber\\
&\times& \left[e^{-\frac{im}{2\hbar t}
\frac{\sigma_0}{s_t}[z-\frac{\hbar t}{m} k_0 + \frac{gt^2}{2}]^2}
+\frac{\alpha}{2i} e^{-\frac{im}{2\hbar t}
\frac{\sigma_0}{s_t}[z-\frac{\hbar t}{m}(k_0+\beta) +
\frac{gt^2}{2}]^2} - \frac{\alpha}{2i} e^{-\frac{im}{2\hbar t}
\frac{\sigma_0}{s_t}[z-\frac{\hbar t}{m}(k_0-\beta) +
\frac{gt^2}{2}]^2} \right]
\end{eqnarray}
where,
%\begin{eqnarray*}
$s_t = \sigma_0 \left( 1+i\frac{\hbar t}{2 m \sigma_0^2} \right)$ and $\beta = \frac{\pi}{4\sigma_0}$.
%\end{eqnarray*}
The probability density is given by  
\begin{eqnarray}
\rho(z, t) &=& |\psi(z, t)|^2 = N^2 \frac{\sigma_0}{\sigma_G} e^{E_1} \left[ e^{E_2} + \alpha \left( e^{E_3} \sin{A_1} + e^{E_4} \sin{A_2} \right) 
+ \frac{\alpha^2}{4} \left( 1 + e^{E_5} -2 e^{E_6} \cos{A_3} \right) \right]~,
\end{eqnarray}
where
\begin{eqnarray*}
E_1 &=& -\frac{\left[ \pi \hbar t + 4m\sigma_0 \left(z- u t +\frac{1}{2}gt^2 \right) \right]^2}{32 m^2 \sigma_G^2 \sigma_0^2}~,\\
E_2 &=& \frac{ \pi \hbar t \left[ \pi \hbar t + 8m\sigma_0 \left(z - u t + \frac{1}{2}gt^2 \right) \right]}{32 m^2 \sigma_G^2 \sigma_0^2}~,\\
E_3 &=& \frac{\pi \hbar t \left[ \pi \hbar t + 24m\sigma_0 \left( z - u t + \frac{1}{2}gt^2 \right) \right]}{64 m^2 \sigma_G^2 \sigma_0^2}~,\\
E_4 &=&\frac{\pi \hbar t \left[ \pi \hbar t + 8 m\sigma_0 \left(z- u t + \frac{1}{2}gt^2 \right) \right]}{64 m^2 \sigma_G^2 \sigma_0^2} ~,\\
E_5 &=& \frac{ \pi \hbar t \left( z-ut +\frac{1}{2}gt^2 \right)}{2 m \sigma_G^2 \sigma_0}~,\\
E_6 &=& \frac{E_5}{2}~,\\
A_1 &=& \frac{\pi \left[- \pi \hbar t + 8 m \sigma_0 \left( z-ut +\frac{1}{2}gt^2 \right) \right]}{32 m \sigma_G^2}~,\\
A_2 &=& \frac{\pi \left[\pi \hbar t + 8 m \sigma_0 \left(z-ut +\frac{1}{2}gt^2 \right) \right]}{32 m \sigma_G^2}~,\\
A_3 &=& \frac{\pi\sigma_0 \left(z-ut +\frac{1}{2}gt^2 \right)}{2 \sigma_G^2}~,
\end{eqnarray*}
and
\begin{eqnarray}
\sigma_{G} &=& |s_t| = \sigma_{0}\left(1+\frac{\hbar^{2}t^{2}}{4m^{2}\sigma_{0}^{4}}\right)^{\frac{1}{2}}~,
\end{eqnarray}
is the spreading of the Gaussian wave packet.
The probability density is explicitly mass dependent. Here, the spreading of the wave packet is given by

\begin{eqnarray}
\sigma_{NG} & = & \frac{\sqrt{\lambda^{(0)} + \lambda^{(2)} \alpha ^2 + \lambda^{(4)} \alpha ^4 }}
{4 m \sigma_{0} \left[ 2e^{\frac{\pi^{2}}{8}} + \alpha^{2} \left( e^{\frac{\pi^{2}}{8}}-1 \right) \right]}
\end{eqnarray}
where,
\begin{eqnarray}
\lambda^{(0)} &=& 64 e^{\frac{\pi^{2}}{4}}m^{2}\sigma_{0}^{2}\sigma_{G}^{2}  ~,\nonumber \\
\lambda^{(2)} &=& 8 e^{\frac{\pi^{2}}{8}}\pi^{2}m^{2}\sigma_{0}^{4} \left(1 - 2 e^{\frac{\pi^{2}}{16}}  \right) 
+64 e^{\frac{\pi^{2}}{8}}m^{2}\sigma_{0}^{2}\sigma_{G}^{2} \left(-1 + e^{\frac{\pi^{2}}{8}} \right)
+2 e^{\frac{\pi^{2}}{4}}\pi^{2}\hbar^{2}t^{2}  ~,\nonumber \\
\lambda^{(4)} &=& \left(e^{\frac{\pi^{2}}{8}} -1 \right) \left[ 16 m^{2}\sigma_{0}^{2}\sigma_{G}^{2} \left(e^{\frac{\pi^{2}}{8}} -1 \right)
+ 4 \pi^{2}m^{2}\sigma_{0}^{4} + e^{\frac{\pi^{2}}{8}}\pi^{2}\hbar^{2}t^{2} \right]~.
\end{eqnarray}
which is also mass dependent. 

\begin{figure}
\centering
\includegraphics[width=12cm,angle=0]{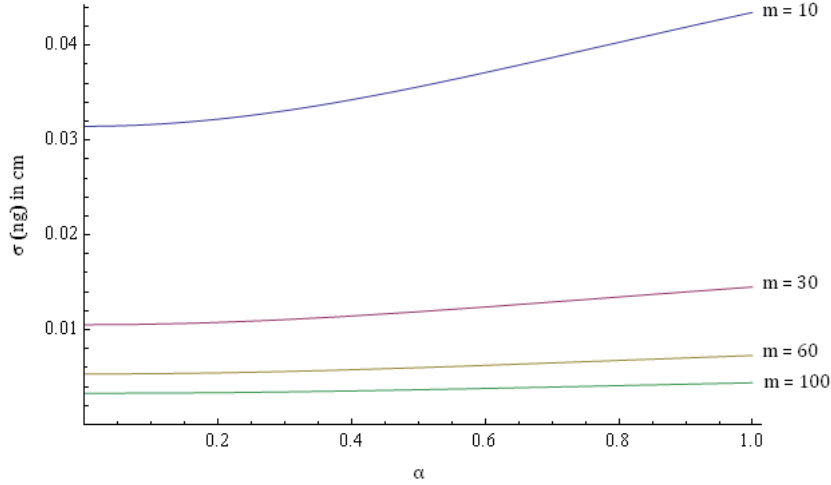}
\caption{The variation of the width of the wave packet with $ \alpha $ for different constant values of $ m $ (in a.m.u.). We take $ \sigma_{0}=10^{-3} $ cm, $ t=1 $ sec and $ u=10^{3} $ cm $ s^{-1} $.}
\vspace*{1cm}
\label{fig: sigma-vs-alpha}
\end{figure}

Fig. \ref{fig: sigma-vs-alpha} gives the plot of $ \sigma_{NG} $ vs $ \alpha $.

%--------------------------- Begin Figure 1      ---------------------------

%--------------------------- End Figure 1      ---------------------------

From this figure, we can see that for large masses, $ \sigma_{NG} $ becomes almost constant with $ \alpha $. For a particular value of $ \alpha $, $ \sigma_{NG} $ decreases with increasing value of mass $ m $.\\

The expression for expectation value of z is given by
\begin{eqnarray} \label{eq: z_peak}
\langle z\rangle = \frac{\pi\alpha\sigma_{0}}{2+\alpha^{2}\left( 1-e^{-\frac{\pi^{2}}{8}}\right)}\,e^{-\frac{\pi^{2}}{32}}+ut-\frac{1}{2}gt^{2}
\end{eqnarray}

Ehrenfest's theorem states that expectation value of quantum mechanical operators obey classical laws of motion, the instances of which are given below
\begin{eqnarray}
\frac{d\langle z\rangle}{dt}=\frac{\langle p\rangle}{m},\nonumber\\
\frac{d\langle p\rangle}{dt}=\langle -\frac{\partial V}{\partial z}\rangle.
\end{eqnarray}
It is possible to check that Ehrenfest's theorem is satisfied for our non-Gaussian wave packet. In the Gaussian limit (i.e., $ \alpha \rightarrow 0 $), $ \langle z\rangle = z_{peak} $ and the dynamics of $\langle z\rangle$ is like a classical point particle, as expected.\\

In Eq. (\ref{eq: z_peak}) the first term is $ \alpha $-dependent. There is no $m$ in this expression. So, $ \langle z\rangle $ is $ \alpha $-dependent but mass independent. With the solutions obtained from the equation $ \frac{d|\psi|^{2}}{dz}=0 $, we can get the value of $ z_{peak} $ for which $ |\psi|^{2} $ is maximum. But the equation can't be solved analytically. So, we have computed $ z_{peak} $ numerically. $ z_{peak} $ depends both on $ \alpha $ and mass. In Table 1, it is shown numerically how $ z_{peak} $ and $ \langle z\rangle $ vary with $ \alpha $, for $0 \leq \alpha \leq 1$. In Table 2, it is shown numerically how those two vary with mass. It is clear that $ z_{peak}\neq\langle z\rangle $ and $ z_{peak} $ increases with $ \alpha $ and mass $ m $ but $ \langle z\rangle $ increases with $\alpha$,  and remains constant for all masses.\\

\begin{figure}
\centering
\includegraphics[width=12cm,angle=0]{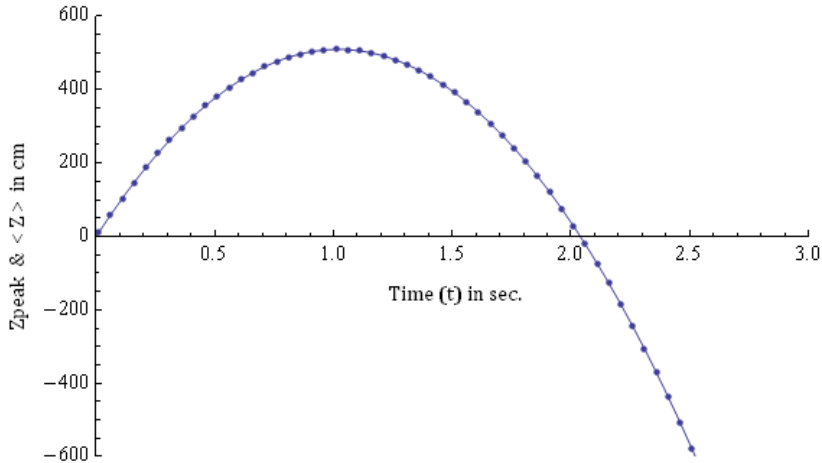}%
\caption{The variation of $ z_{peak} $ and $ \langle z\rangle $ with time $ t $. We take $ \sigma_{0}=10^{-7} $ cm, $ u=10^{3} $ cm $ s^{-1} $, $m=50$ a.m.u. and $\alpha=0.5$.}
%\vspace*{.8cm}
\label{fig: z-vs-t}
\end{figure}

Fig. \ref{fig: z-vs-t} shows the variation of $ z_{peak} $ and $ \langle z\rangle $ with time simultaneously. Dotted curve shows the motion of $ z_{peak} $ whereas the continuous curve shows the motion of $ \langle z\rangle $. As we take $ \sigma_{0}=10^{-7} $ cm, the $ \alpha $-dependence of $ \langle z\rangle $ and $ z_{peak} $ becomes vanishingly small. So, this $ \alpha $-dependence can not be shown graphically. As a result, the curves for $ z_{peak} $ and $ \langle z\rangle $ coincide totally in this graph.

%--------------------------- Begin Figure 2      ---------------------------

%--------------------------- End Figure 2      ---------------------------

\begin{table}
\begin{flushleft}
\caption{The variation of $ \langle z\rangle $ and $ z_{peak} $ in cm with $ \alpha $ for mass $ m=10 $ a.m.u. $ t=2 $ sec and $ \sigma_{0}=0.1 $ cm.}
\end{flushleft}
\begin{center}
\begin{tabular}{|ccc|}
\hline $ \alpha $ & $ z_{peak} $ & $ \langle z\rangle $\\
& in cm & in cm\\
\hline
\hline 0 & 38.59999999999815 & 38.599999999999910 \\
\hline 0.1 & 38.61540633397262 & 38.611498366598200 \\
\hline 0.2 & 38.62925626009014 & 38.622755653868810 \\
\hline 0.3 & 38.64090000000000 & 38.633500000000000 \\
\hline 0.4 & 38.65026513575232 & 38.643679695976970 \\
\hline 0.5 & 38.65786585282045 & 38.652999876917650 \\
\hline 0.6 & 38.66400000000000 & 38.661400000000000 \\
\hline 0.7 & 38.66910000000000 & 38.668800000000000 \\
\hline 0.8 & 38.67337906376510 & 38.675246198243485 \\
\hline 0.9 & 38.67696171641449 & 38.680689435211890 \\
\hline 1.0 & 38.68002216630916 & 38.685197660661515 \\
\hline
\end{tabular}
\end{center}
\end{table}

\begin{table}
\begin{flushleft}
\caption{The variation of $ \langle z\rangle $ and $ z_{peak} $ in cm with mass $ m $ in a.m.u. for $ \alpha=0.5 $, $ t=2 $ sec and $ \sigma_{0}=0.1 $ cm.}
\end{flushleft}
\begin{center}
\begin{tabular}{|ccc|}
\hline Mass $ (m) $ & $ z_{peak} $ & $ \langle z\rangle $ \\
in a.m.u. & in cm & in cm\\
\hline
\hline 30 & 38.657865898827980 & 38.65299987691765 \\
\hline 60 & 38.657865903136155 & 38.65299987691765 \\
\hline 90 & 38.657865903934166 & 38.65299987691765 \\
\hline 120 & 38.657865904213070 & 38.65299987691765 \\
\hline 150 & 38.657865904342295 & 38.65299987691765 \\
\hline
\end{tabular}
\end{center}
\end{table}

Now, we want to observe the expression for $ \langle z\rangle $ in the classical limit. If a full classical description is to emerge from quantum mechanics, we must be able to describe quantum systems in phase-space. The Wigner distribution function \cite{Wi-PR-1932, FoCo-AJP-2002} which is one of the phase-space distributions, can provide a re-expression of quantum mechanics in terms of classical concepts.\\

The Wigner distribution function which is calculated from the time evolved wave function $ \psi(z,t) $, is given by
\begin{equation}
D_{w}(z,p,t)=\frac{1}{\pi\hbar}\int_{-\infty}^{\infty}\psi^{*}(z+y,t)\psi(z-y,t) e^{\frac{2ipy}{\hbar}}dy
\end{equation}
and the marginals of which yield the correct quantum probabilities for position and momentum separately.\\

By substituting the value of $ \psi^{*}(z+y,t) $ and $ \psi(z-y,t) $ using Eq.(5) we obtain the Wigner distribution function $ D_{w}(z,p,t) $ for non-Gaussian wave packet. It satisfies the classical Liouville's equation \cite{DaSe-CS-2002} given by
\begin{equation}
\frac{\partial D_{w}(z,p,t)}{\partial t}+\dot{z}\frac{\partial D_{w}(z,p,t)}{\partial z}+\dot{p}\frac{\partial D_{w}(z,p,t)}{\partial p}=0
\end{equation}
Here, $ \dot{z}=p/m $ and $ \dot{p}=-mg $ are obtained from Hamilton's equation of motion given by
\begin{eqnarray}
\dot{z}=\frac{\partial H}{\partial p},\nonumber\\
\dot{p}=-\frac{\partial H}{\partial z}\nonumber\\
\end{eqnarray}
where, $ H=p^{2}/2m + mgz $ is the Hamiltonian of the system.\\

The corresponding position distribution function is given by
\begin{equation}
\rho_{C}(z,t)=\int_{-\infty}^{\infty}D_{w}(z,p,t)dp.
\end{equation}
So, the expression for the average value of $ z $ using Wigner distribution function is given by
\begin{eqnarray}
\langle z\rangle & = & \frac{\int_{-\infty}^{\infty}z\rho_{C}(z,t)dz}{\int_{-\infty}^{\infty}\rho_{C}(z,t)dz}\nonumber\\
 & = & \frac{\pi\alpha\sigma_{0}}{2+\alpha^{2}\left( 1-e^{-\frac{\pi^{2}}{8}}\right)}\,e^{-\frac{\pi^{2}}{32}}+ut-\frac{1}{2}gt^{2} \equiv z_0 + ut-\frac{1}{2}gt^{2}
\end{eqnarray}
which is exactly equal to the quantum mechanical expectation value of $ z $.\\

The particles are projected upwards against gravity. Cassicaly, the refrence point $z=0$ reaches the maximum height $z=z_{max}=ut_{1}-\frac{1}{2}gt_{1}^{2} $ at the time $ t=t_{1}=u/g $. So, the probability of finding the particles $ P_{1}(m) $ within a very narrow detector region $ (-\:\epsilon\: to\: +\:\epsilon) $ around $ z_{max} $ is given by\\
\begin{equation}
P_{1}(m)=\int_{z_{max}-\epsilon}^{z_{max}+\epsilon}|\psi(z,t_{1})|^{2}dz.
\end{equation}\\

At a later time $ t=t_{2}=2u/g $, the classical peak of the wave packet returns to it's initial projection point at $ z=0 $. Similarly, the probability of finding the particles $ P_{2}(m) $ within a very narrow detector region $ (-\:\epsilon\: to\: +\:\epsilon) $ around $ z=0 $ is given by\\
\begin{equation}
P_{2}(m)=\int_{-\epsilon}^{+\epsilon}|\psi(z,t_{2})|^{2}dz.
\end{equation}\\

\begin{figure}
\centering
\includegraphics[width=10cm,angle=0]{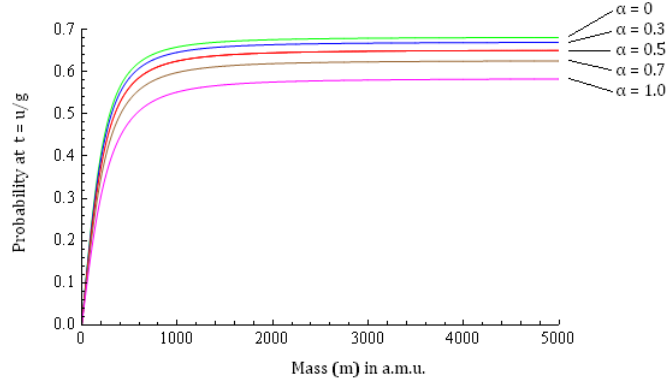}%
\caption{The variation of probability $P_1(m)$ with mass (in a.m.u.) for different constant values of $ \alpha $. We take $ u=10^{3} $ cm $ s^{-1} $, $ \sigma_{0}=10^{-3} $ cm and $ \epsilon=\sigma_{0} $.}
%\vspace*{.8cm}
\label{fig: pd-vs-t1}
\end{figure}

\begin{figure}
\centering
\includegraphics[width=10cm,angle=0]{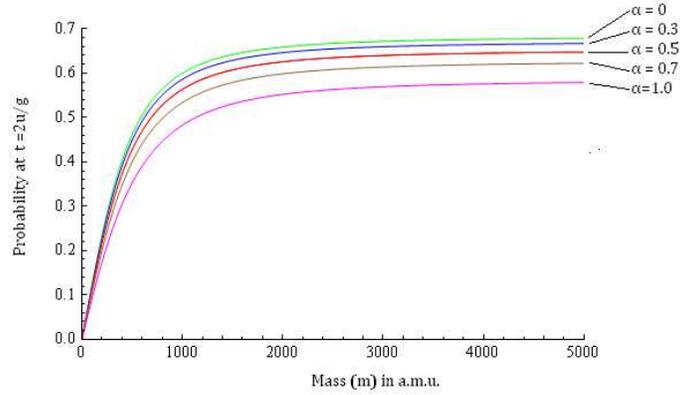}%
\caption{The variation of probability $P_2(m)$ with mass (in a.m.u.) for different constant values of $ \alpha $. We take $ u=10^{3} $ cm $ s^{-1} $, $ \sigma_{0}=10^{-3} $ cm and $ \epsilon=\sigma_{0} $.}
%\vspace*{.8cm}
\label{fig: pd-vs-t2}
\end{figure}

From Fig. \ref{fig: pd-vs-t1} and Fig. \ref{fig: pd-vs-t2}, it is clear that both the probabilities are mass dependent for an initial non-Gaussian position distribution for smaller masses. This effect of mass dependence of the probabilities occur essentially as the spreading of the wave packet under linear gravitational potential depends on the mass of the particles. These probabilities become saturated in the limit of large mass. As $ \alpha $ increases, both the probabilities decrease for a particular value of $m$. The departure from the result
for the Gaussian ($\alpha = 0$) case increases with the value of the non-Gaussian parameter, signifying a stronger violation of the weak equivalence principle.

%--------------------------- Begin Figure 3      ---------------------------

%--------------------------- End Figure 3      ---------------------------

%--------------------------- Begin Figure 4      ---------------------------

%--------------------------- End Figure 4     ---------------------------

%NewSection-NewSection-NewSection-NewSection-NewSection-NewSection-NewSection-

\section{Mean Arrival Time}
\vspace{1cm}
\hspace{0.5cm}

Now, Let us consider the quantum particles subjected to free fall from the initial position at $ z=0 $ under gravity with the initial state given by Eq.(1) and with $ u=0 $. Using the probability current approach \cite{prob} the mean arrival time of the quantum particles to reach a detector location at $ z=Z $ is given by
\begin{equation}
\bar{\tau}=\frac{\int_{0}^{\infty}|J(z=Z,t)|\,t\, dt}{\int_{0}^{\infty}|J(z=Z,t)|\,dt}\\
\label{tau}
\end{equation}
where, $ J(z,t) $ is the quantum probability current density. However, we emphasize that the definition of the mean arrival time used in the above equation is not a uniquely derivable result within standard quantum mechanics. It should also be noted that $ J(z,t) $ can be negative, so we take the modulus sign in order to use the above definition. From Eq.(\ref{tau}), it can be seen that the integral of the denominator converges whereas the integral of the numerator diverges formally. To avoid this problem, several techniques have been employed \cite{DaEgMu-quant-2001}. In this paper, to obtain convergent results of the numerator we use the simple technique of getting a cut-off at $ t=T $ in the upper limit of the time integral with $ T=\sqrt{2(| Z |+3\sigma_{T})/g} $. $ \sigma_{T} $ is the width of the wave packet at time $ T $. Thus, our calculations of the arrival time are valid upto the $ 3\sigma_{T} $ level of spread in the wave function. \\

Here, the expression for the Schr\"{o}dinger probability current density $ J(z,t) $ with the initial non-Gaussian position distribution is given by
\begin{eqnarray}
J(z,t )&=& \frac{\hbar}{m} \Im{(\psi^* \nabla \psi)} = N^2 \frac{e^{E_1}}{32 m^2 \sigma_0 \sigma_G^3} \left( \eta^{(0)}  + \alpha \eta^{(1)} + \alpha^2 \eta^{(2)} \right)~,
\end{eqnarray}
where
\begin{eqnarray*}
\eta^{(0)} &=& 8 e^{E_2} \left[ \hbar^2 t \left( z-\frac{1}{2} gt^2 \right) + 4m^2\sigma_0^4 \left( u -gt \right)  \right]~,\nonumber\\
\eta^{(1)} &=& \eta^{(0)} e^{-E_2} \left( e^{E_3} \sin{A_1} + e^{E_4} \sin{A_2} \right) -2\pi \hbar^2 t \sigma_0 \left( e^{E_3} \cos{A_1} + e^{E_4} \cos{A_2} \right) + 4\pi m \hbar \sigma_0^3 \left( e^{E_3} \sin{A_1} - e^{E_4} \sin{A_2} \right) ~,\nonumber\\
\eta^{(2)} &=& \left[ 2 \hbar^2 t \left(  z-\frac{1}{2} gt^2 \right) + 8 m^2 \sigma_0^4 \left( u -gt \right) \right]
(1 + e^{E_5} - 2 e^{E_6} \cos{A_3}) -2 \pi \hbar^2 t \sigma_0 e^{E_6} \sin{A_3} + 2\pi m \hbar \sigma_0^3 (-1 + e^{E_5})~,
\end{eqnarray*}
and where the $E_i$'s are defined below Eq.(7).

\begin{figure}
\centering
\includegraphics[width=10cm,angle=0]{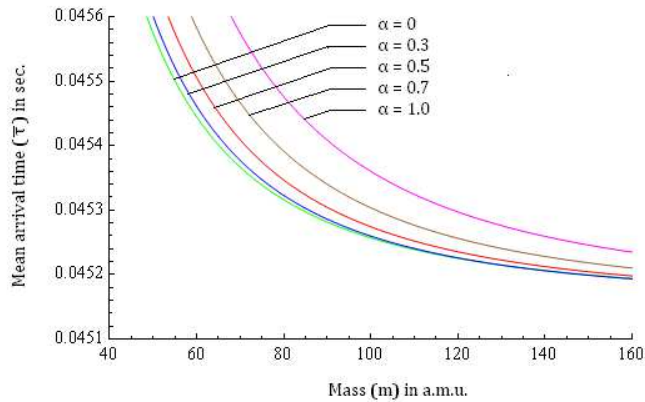}%
\caption{The variation of mean arrival time with mass (in a.m.u.) for different constant values of $ \alpha $. We take $ \sigma_{0}=10^{-6} $ cm, $ Z=-1 $cm and $u=0$.}
%\vspace*{.8cm}
\label{fig: tau-vs-mass}
\end{figure}

It is clear that $ J(z,t) $ is explicitly mass dependent. So, $ \bar{\tau} $ is also mass dependent. From Fig. \ref{fig: tau-vs-mass}, we can observe that for smaller masses $ \bar{\tau} $ is mass dependent and for larger masses it becomes mass independent. For a particular value of mass, $ \bar{\tau} $ increases with $ \alpha $.

%--------------------------- Begin Figure 5     ---------------------------

%--------------------------- End Figure 5    ---------------------------

\begin{figure}
\centering
\includegraphics[width=10cm,angle=0]{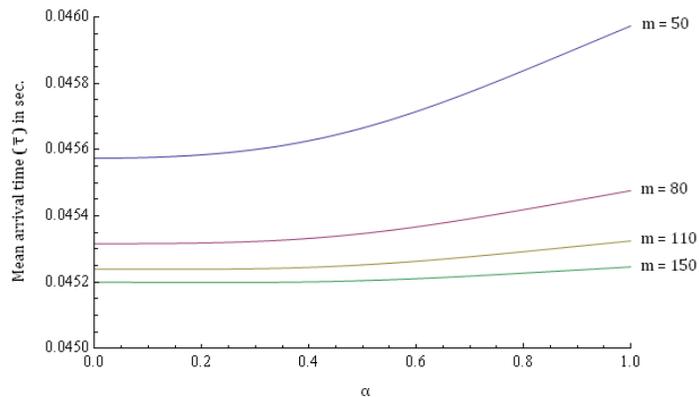}%
\caption{ The variation of mean arrival time with $ \alpha $ for different constant values of $ m $ (in a.m.u.).We take $ \sigma_{0}=10^{-6} $ cm, $ Z=-1 $cm and $u=0$.}
%\vspace*{.8cm}
\label{fig: tau-vs-alpha}
\end{figure}

From Fig. \ref{fig: tau-vs-alpha}, we can see that for large masses, $ \bar{\tau} $ becomes almost constant with $ \alpha $. For a particular value of $ \alpha $, $ \bar{\tau} $ decreases with increasing value of mass $ m $.

%--------------------------- Begin Figure 6     ---------------------------

%--------------------------- End Figure 6    ---------------------------

It is clear that the mass-dependence of the mean arrival time gradually
vanishes as the mass is increased. Thus, compatibility with the equivalence
principle emerges in the limit of large mass, or classical limit. Note
that with the increase of the parameter $\alpha$, one needs to go to
higher values of mass in order for the mass-dependence to become insignificant.
Thus, a non-Gaussian wave packet can exhibit non-classical features in a 
mass range where its Gaussian counterpart behaves classically.

%NewSection-NewSection-NewSection-NewSection-NewSection-NewSection-NewSection-

\section{Bohm Trajectory Approach}
\vspace{1cm}
\hspace{0.5cm}

Within the Bohmian interpretation of quantum mechanics \cite{Bo-PR-1952}  each individual particle is assumed to have a definite position, irrespective of any measurement. The equation of motion of the particle is,
\begin{eqnarray}
m \ddot{\vec{x}} &=& -\nabla(V + Q) |_{\vec{x}=\vec{x}(t)}~,
\end{eqnarray}
where $Q = -\frac{\hbar^2}{2m} \frac{\nabla^2 \sqrt{\rho}}{\sqrt{\rho}}$ is called 'quantum potential energy' and it is explicitly mass-dependent and $\vec{x}(t)$ determines the path of the particle. In the gravitational potential $V = mgz$ the quantum version of Newton's second law is given by
\begin{eqnarray}
\ddot{z} &=& g - \frac{1}{m} \frac{\partial Q}{\partial z} \bigg|_{z=z(t)}~.
\end{eqnarray}
To quote from  Holland \cite{Ho-book-1993}, "It is immediately clear that the assumption of the equality of the inertial and passive gravitational masses does not imply that all bodies fall with an equal acceleration in a given gravitational field, due to the intervention of the mass-dependent quantum force term." Thus,  WEQ is violated.  The arrival-time problem is unambiguously solved
in the Bohmian mechanics, where for an arbitrary scattering potential $V(\vec{x})$, one finds \cite{Le-book-2002}
that for those particles that actually reach $\vec{x} = \vec{X}$, the arrival-time distribution is given by the
modulus of the probability current density, i.e., $|J(\vec{X}, T )|$. 

\begin{figure}
\centering
\includegraphics[width=10cm,angle=-90]{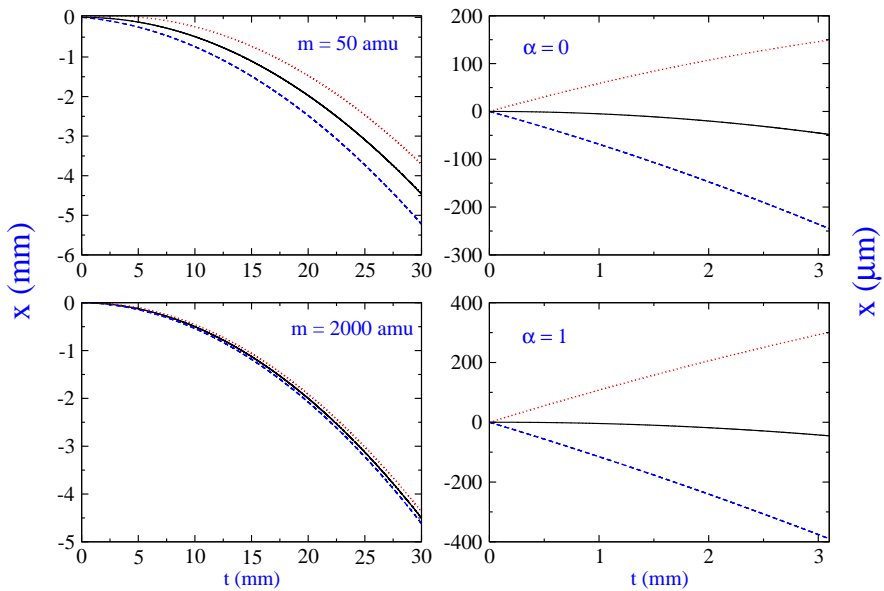}
\caption{A selection of Bohmian trajectories for $u=0$ (free fall) and $\sigma_0 = 10^{-6}$cm. In the left column tunable parameter $\alpha$ has the fixed value $\alpha = 0.5$ and in the right one mass is constant and equal to $m = 100~$amu. The solid black trajectory starts at $\langle z \rangle(0)$ , the dashed blue one at $\langle z \rangle(0) - 2 \sigma_0$ and the dotted red one at $\langle z \rangle(0) + 2 \sigma_0$.}
%\vspace*{.8cm}
\label{fig: freefall-traj}
\end{figure}

The quantum potential $Q$ is nonlocal, and thus it is expected that its
effect on the violation of WEQ is manifested in a way similar to the violation
observed as a consequence of the spread of the wave packet. We compute a 
set of Bohmian trajectories corresponding to free fall of our non-Gaussian
wave packet with the tunable parameter $\alpha$.
Fig. \ref{fig: freefall-traj} shows a selection of Bohmian trajectories
exhibiting their spread depending upon the values of $m$ and $\alpha$.
Note that for small mass, the trajectories with initial positions on the
left and right of the center of the wave packet (mean $\langle z\rangle$)
spread out with evolution, indicating violation of WEQ. The magnitude
of spread increases with $\alpha$, signifying stronger violation of WEQ
with increased departure from Gaussianity. The spread of the trajectories 
decreases as mass is increased, leading to the emergence of WEQ in
the classical limit.

%-------------------- Begin Fig7 -------------------

%-------------------- End Fig7----------------------
 
%NewSection-NewSection-NewSection-NewSection-NewSection-NewSection-NewSection-
\section{Conclusions}
\vspace{1cm}
\hspace{0.5cm}

To summarize, in this work we have studied the violation of the gravitational
weak equivalence principle in quantum mechanics using a non-Gaussian wave
packet. The wave packet is constructed in such a way, that its departure
from Gaussianity is represented by a continuous and tunable parameter $\alpha$.
The asymmetry of the wave packet entails a difference between its mean and peak,
and causes the peak to evolve differently from a classical point particle, whereas
the mean evolves like a classical point particle modulo a constant depending upon
the parameter $\alpha$.
Such a result is consistent with the Ehrenfest therorem, as we have shown,
and re-confirmed using the Wigner distribution. The correspondence with
the results following from a Gaussian wave packet is achieved in the limit
of $\alpha \to 0$.

The weak equivalence principle is shown to be violated first through the 
dependence on
mass of the position detection probabilities of freely falling particles.
The magnitude of violation increases with the increase of the non-Gaussian
parameter $\alpha$, signifying stronger viotation of WEQ with larger
deviation from Gaussianity. We next compute the arrival time of freely
falling wave packets through the probability current distribution 
corresponding to the non-Gaussian wave packet. The violation of WEQ is
again exhibited with the dependence on mass of the mean arrival time.
It is observed that in the limit of large mass the classical value for the
mean arrival time is approached, thereby indicating the emergence of the
WEQ in the classical limit. In this
context it is worthwhile to mention that though our work follows as a
natural consequence of quantum mechanics or quantum theory in the
non-relativistic limit, it does not imply a violation of general covariance
in the relativistic domain. An interesting connection between the
non-relativistic limit of quantum theory and the principle of equivalence
has recently been discussed \cite{PaPa-PRD-2011}. Finally, a computation of Bohmian trajectories further
establishes the stronger violation of WEQ by a non-Gaussian wave packet.
We conclude by noting that realization of exactly Gaussian wave packets is
rather difficult in real experiments. One of the consequences of our results 
is to enable relaxation of the wave packet to be Gaussian, and therefore it
can facilitate the experimental observation of the violation of WEQ, as
well as enable a quantitative verification of the way the violation of WEQ
depends on the departure from the Gaussian nature of the wave packet. Other
ramifications of this result call for further study.  \\

{\it Acknowledgements:} PC, DH, ASM and SS acknowledge support from DST India
Project no. SR/S2/PU-16/2007. MRM and SVM acknowledge finantial support of the
University of Qom.

\end{document}